\documentclass[12pt]{article}
\author{Zosia A. C. Krusberg\footnote{Electronic address: zosia@uchicago.edu}\\
\\
\small \it Department of Physics, University of Chicago\\
\small \it 5640 S. Ellis Avenue, Chicago, IL 60637, USA}
\title{Emerging technologies in physics education}
\date{}
\begin{document}
\maketitle

\begin{abstract}

\noindent Three emerging technologies in physics education are evaluated from the interdisciplinary perspective of cognitive science and physics education research.  The technologies --- Physlet Physics, the Andes Intelligent Tutoring System (ITS), and Microcomputer-Based Laboratory (MBL) Tools --- are assessed particularly in terms of their potential at promoting conceptual change, developing expert-like problem-solving skills, and achieving the goals of the traditional physics laboratory.  Pedagogical methods to maximize the potential of each educational technology are suggested.

\end{abstract}

\section{Introduction}

The use of educational technologies in university physics courses has increased dramatically in the last decade.  Course management software is placing course documents, resources, and student grades online.  Homework delivery systems are supplying physics instructors with databases of potential homework problems that students can complete online.  Electronic response systems and simulations are transforming the traditional physics lecture into an interactive and collaborative learning environment.  Elaborate software is teaching students how to problem solve and to perform sophisticated mathematical calculations.

Despite the dramatic changes caused by the implementation of technology to many aspects of university physics courses, these technologies have not offered a panacea to the ever-deepening problems in contemporary physics education.  Though over the past few decades, cognitive science and physics education research have provided deep insight into how students at all educational levels learn physics, instructional methods in most university physics courses do not differ significantly from those of a century ago.  Fortunately, the increasing sophistication and availability of educational technologies is providing physics educators with an opportunity to reevaluate the goals of physics instruction, and, armed with the growing body of research in cognitive science and physics education, use educational technologies to achieve these goals.

Thereby, the goal of this paper is to assess three emerging technologies in physics education from the interdisciplinary perspective of cognitive science and physics education research.  The three technologies address different aspects of physics knowledge: Physlets, a collection of Java applets, are designed to deepen students' conceptual knowledge of physics; Andes Intelligent Tutoring System (ITS) aims to develop students' procedural knowledge of physics problem solving; and Microcomputer-Based Laboratory (MBL) Tools, a set of probes and associated software, seek to relieve the physics laboratory of the drudgery of data collection and display.  The subsequent sections will present the technical specifications of each technology and the central objectives guiding its design, and discuss how the technology offers an improvement over existing instructional methods and how it aligns with findings in cognitive science and physics education research.  Finally, some thoughts about the future outlook of the technology will be presented.

\section{Physlet Physics}

\subsection{Technical Specifications}

Physlets, developed by Wolfgang Christian and Mario Belloni at Davidson College, are small, flexible, Java-generated computer animations that can be embedded into html documents and run on nearly every web platform.  The animations generally provide visualizations of multiple representations of a specific physical phenomenon, such as simulations, graphs, diagrams, and tables.  A number of control buttons below the applet itself allow students to start, stop, and step the animations, and the mouse can be used to read scaled coordinates and to drag and drop objects around the frame.  Each Physlet is designed to focus on a single physical principle or concept, excluding unnecessary detail; this keeps Physlets small and easily downloadable over the internet on a range of connection speeds.  Also, as a result of their simplicity, Physlets do not require instructors to adhere to a particular pedagogical approach, though the creators point out that Physlets are most effective when utilized in collaborative learning or tutorial-type settings.  A large collection of Physlets is available for free download from the Physlet website, and is accompanied by extensive digital resources as well as a number of books and journal articles.  The Physlet exercises offer nearly complete coverage of the introductory university physics sequence, and, more recently, have extended their coverage into introductory quantum mechanics \cite{cn2005a,cn2005b}.  

Recently, the Physlet creators published \emph{Physlet Physics}\cite{cb2004a}, a book containing over 800 applets and whose content span the full introductory physics sequence; this publication seeks to provide physics instructors with a complete and structured collection of Physlets that can be implemented into existing physics curricula.  Every chapter in Physlet Physics contains three different types of Physlet exercises --- Illustrations, Explorations, and Problems --- intended to be completed in a specific sequence, and each utilizing a slightly different approach to help students develop an understanding of various physics concepts.  The book, which includes a CD-ROM containing the Physlet collection, provides snapshots of the applets as well as full versions of the text and questions that appear along with the online Physlets.  

Each chapter of Physlet Physics opens with a number of Illustrations seeking to introduce students to new physical concepts.  The chapter Refraction (Chapter 34), for instance, opens with three Illustrations: Huygens' Principle and Refraction (34.1), Fiber Optics (34.2), and Prisms and Dispersions (34.3).  A brief text accompanying each applet briefly explains the underlying physical principle; the text accompanying Illustration 34.3 states:  ''The index of refraction of a given material depends on the wavelength (or frequency) of the incoming light.  Hence, the speed of light in that material also depends on the wavelength of frequency of light'' \cite{cb2004b}.  In general, Illustrations are interactive; the Refraction Illustrations, for instance, allow students to change the index of refraction of a material, alter the angle of incidence of light in a fiber-optic cable, and modify the wavelength of a light beam incident on a prism.  If the Illustration poses a question to the student, the answer is either provided directly in the text or indirectly from interaction with the applet.  In a curriculum that has integrated Physlets into its instruction, Illustrations can be used to introduce new physical concepts or analytical tools, either in the form of homework assignments or in classroom demonstrations.  

Explorations, on the other hand, serve as tutorials, guiding students' interactions with the applet and providing hints and suggestions to conceptual and procedural problems.  Many Explorations ask students to make predictions prior to observing a particular physical phenomenon.  In Exploration 34.1, for example, students are asked to predict how light emerging from a convex lens would behave if the index of refraction of the outside material were increased; then, students can vary the index of refraction using a slider located beneath the animation and test their predictions.  Other Explorations let students alter the values of different physical parameters in order to establish or verify physical relationships.  In the Refraction chapter, for example, students are asked to verify the validity of Snell's Law with the help of a virtual pink protractor (Exploration 34.2), and later, the relationship between the curvature of a mirror and its focal length (Exploration 34.3).  The CD-ROM accompanying Physlet Physics also contains supplemental Exploration Worksheets designed to provide students with additional scaffolding when completing the Explorations; these worksheets contain specific spaces in which students can write or draw predictions or answers to questions.  If implemented into the regular curriculum, Explorations function well as group problem-solving challenges or pre-laboratory assignments.  

Finally, Problems resemble Explorations in that both provide students with exercises that require varying degrees of conceptual or procedural understanding; however, unlike the Explorations, Problems provide students with little guidance or scaffolding.  Problems are most effectively used as homework assignments, group problem-solving challenges, or as classroom discussion questions.  

\subsection{Analysis}

The principal objective of the Physlet collection is to develop and deepen students' conceptual understanding of fundamental physics principles.  Fortunately, the study of how children learn scientific concepts --- particularly in physics --- has been an active area of research both in cognitive science and science education research for several decades \cite{ps1941,cs1994,gro2003}.  One of the most important findings of this research area is that students enter the classroom not as blank slates, but with firmly rooted conceptions of the physical world.  Unfortunately, these so-called preconceptions are often incorrect by formal scientific standards.  In the last decade, volumes have been dedicated to documenting the most common physics preconceptions, in order for physics instructors to be aware of the initial cognitive states of their students, and to allow them to adapt their curricula accordingly \cite{aro1997,maz1997,kni2004}.  

Based on these findings, science education researchers have established instructional methods and curricula designed specifically to modify students' preconceptions to align with formal scientific theories.  These methods --- generally based on the so-called cognitive conflict principle --- tend to incorporate the following critical components: first, students' preconceptions need to be elicited, either by having students make a prediction to the outcome of a demonstration, or answer a conceptual question.  Vast collections of these demonstrations and questions exist, designed based on research findings specifically to bring out common misconceptions \cite{maz1997}.  Then, students are presented with a scenario --- such as a demonstration, a simulation, or an additional problem --- where their preconception fails to explain a given phenomenon.  The new, scientifically correct conception is then introduced, and shown to succeed where the preconception failed.  Finally, students are provided with opportunities to practice utilizing and solidifying the new conception \cite{gro2004}.

The pedagogical structure of the Physlets adheres very well to this model for inducing conceptual change.  After a brief introduction to a given physical principle via the Illustrations, the Explorations can serve to bring out students' misconceptions about this principle by means of a prediction regarding the outcome of a simulation.  If the students are required to use the Exploration Worksheets, which is recommended, a specific space is provided for students to write or draw their prediction.  In the aforementioned example of light emerging from a convex lens into a medium whose index of refraction the student can alter, for instance, the first question is: ''How, if at all, would the path of the rays change if the source and the lens were placed in another medium with an index of refraction of n = 1.2? [...]  Draw what you expect the rays to look like in the new medium.''  Once the student has made a prediction, the next question is: ''Was your prediction correct?  Explain'' \cite{cb2004a}.  In providing designated spaces for students to make and discuss their predictions, the worksheets thereby scaffold the students' interactions with the applet.  Finally, by supplying students with additional opportunities to practice using the new physical concept in a number of Problems, the Physlets are further adhering to the established conceptual change model. 

Science education researchers have identified another skill as critical to developing a conceptual understanding of scientific principles: the ability to translate across multiple representations.  In science in general, a wide range of different representations are used to represent scientific phenomena, and this is particularly the case in physics.  Physical phenomena can be represented in words, equations, tables of numbers, graphs, and specialized diagrams.  In fact, one of the main differences found between novice and expert physicists is their ability to translate effectively across these representations, and the links between them contribute to experts' coherent knowledge structures and their effective problem-solving strategies.  As will be discussed later in this paper, a number of technologies have been developed specifically to help students solidify these connections between representations \cite{lem2000,red2003}.  

Throughout its simulations, Physlets utilize a variety of these representations in displaying physical phenomena.  Along with its simple visualizations, the applets often include additional frames that display the same phenomenon in a different representation.  In the chapter on Faraday's Law (Chapter 29), for instance, Illustration 29.2: Loop in a Changing Magnetic Field presents the student with an animation of a loop through which the magnetic flux changes; simultaneously, the applet displays a plot of Magnetic Field vs.~Time, and another of Induced Emf vs.~Time.  Visually, then, the applet is able to illustrate the physical relationships inherent in Faraday's Law, namely, that a changing magnetic flux through a loop induces an emf in that loop.  Although no research has explicitly evaluated this aspect of the Physlets, research investigating other technologies that offer students simultaneous representations of physical phenomena have reported great improvement in students' conceptual understanding \cite{ts1990,rss1997,tg2000}.

Finally, the Physlets' tutorial structure aligns well with contemporary theories of social learning that apply to the development of both conceptual and procedural understanding.  These theories were first developed in the late 1970's by Russian psychologist Lev Vygostky \cite{vyg1978}, and have been reinforced more recently by Fischer \& Bidell \cite{fb2005}.  According to these theorists, students' performance at various conceptual and procedural tasks may be greatly enhanced during collaboration with an adult or with more capable peers.  Using Fischer \& Bidell's terminology, when working independently, students perform at a functional level, whereas in collaboration with adults or peers, they perform at the significantly higher optimal level.  This type of scaffolding is provided in the very structure of the Physlets: the instructions and explanations in the Illustrations, as well as the guiding hints and strategies in the Explorations, provide students with the type of support provided by peers or instructors in traditional educational settings.  This scaffolding, then, allows the student to perform at a higher level than possible when facing the problem without support, and the subsequent Problems provide students with the opportunity to solidify their developing skills.  

\subsection{Outlook}

Physlets have enriched physics instruction with an invaluable technological tool, most fundamentally by providing students with dynamic, interactive animations of physical phenomena previously only visualized in static textbook images.  Physics as a discipline, after all, is generally concerned with dynamic phenomena, with static ones representing only special cases of more general principles.  Furthermore, like similar technologies developed for biology and chemistry education \cite{con2006}, Physlets allow students to interact and experiment with highly abstract physical concepts inaccessible in traditional laboratory settings: students can listen to the changing frequency heard from a sound-emitting source whose velocity they can control, move around two point charges to explore the resulting net electric field, and view the motion of a charged particle in a magnetic field.  These general properties of Physlets alone make them an invaluable addition to the physics education toolbox.

However, like any particular educational tool or activity, Physlets may benefit students most when part of a structured curriculum.  As outlined in the previous section, specific instructional methods have been developed to help students modify their scientific misconceptions.  Though the Illustration-Exploration-Problem structure corresponds well with this model --- which is further reinforced and supported by the Exploration Worksheets --- students' adherence to this structure needs monitoring.  Thereby, Physlets may have their greatest impact on students' conceptual learning when supported by a structured curriculum that makes use of collaborative learning under instructor supervision.  Alternatively, Physlets may be unified with intelligent tutoring systems that monitors and provides feedback on student answers to the questions associated with each applet.  

Overall, Physlets are an extremely practical and flexible educational tool.  They are available free over the internet, and are accompanied by extensive online and physically published resources; they are easily modified by physics instructors wanting to adapt them to their specific educational needs; they can be used as introductory material to a new physics concepts, as classroom or homework exercises, and in collaborative groups; and, finally, they are technologically simple enough to be easily integrated into other educational technologies.

\section{Andes Physics Tutoring System}

\subsection{Technical Specifications}

The Andes Physics Tutoring System was developed at the University of Pittsburgh and the United States Naval Academy through collaboration with the Cognitive Science Program at the Office of Naval Research, based on a common interest in developing artificially intelligent tutoring technology for physics education.  To this end, two existing technologies --- Cascade, a rule-based cognitive model of physics problem solving, and Olae, an online assessment system --- were combined and supplemented with the capacity to provide hints and feedback on student work to create Andes.  The fundamental objective of the Andes Intelligent Tutoring System (ITS) was to interact with physics students using the method of coached problem solving, whereby the ITS and the student collaborate through the problem-solving process.  In this process, when the student makes good progress toward a problem solution, the ITS simply agrees with each step; however, if the student makes an error or gets stuck, the ITS can provide hints or feedback on the student error.  Now, Andes is freely available for download, or can be used as a web-based service \cite{van2005,and2006}.  

Four general principles guided the design of Andes.  First, transfer is facilitated by making the ITS interface as similar to a pencil-and-paper solution as possible.  Second, the student is provided with flexibility in the order in which actions are performed, and is allowed to skip steps when applicable.  Third, immediate feedback is provided in order to minimize the amount of time spent pursuing wrong paths toward a solution, and to maximize the opportunities for learning.  Finally, students are encouraged to construct their own knowledge by receiving simple hints that require them to derive most of the problem solution on their own \cite{gv2000}.  

In order to satisfy the first design principle --- to facilitate transfer by making the Andes interface similar to a piece of paper --- the number of structured entry fields in the ITS is minimized.  Thereby, the Andes interface consists of four different panes: two entry panes, one for diagrams and one for equations, one pane for variable definitions, and one for hints.  When a problem is first presented, students are generally called upon to do a qualitative problem analysis by drawing a diagram in the diagram pane.  The area allows for a wide range of different drawings, including free body diagrams, vectors, coordinate systems, angles between vectors and axes, and components of circular paths.  When an object is drawn in this pane, a dialog box is presented that instructs students to define it.  Though some problems consist entirely of the qualitative problem analysis, others go on to require a full algebraic and numerical solution.  Consequently, the next step in the problem-solving process is to define relevant variables.  This aspect of Andes represents its most significant difference from a pencil-and-paper solution, since equations may be included on the latter without having each variable explicitly defined; in Andes, however, in order for a student to enter Newton's Second Law --- $F = m * a$ --- they are first required to define what each of the variables $F$, $m$, and $a$ refer to.  Students can define variables in two ways: either by assigning a variable name to a component of their diagram, or by entering variables in the variable definition menu.  Finally, once relevant variables have been designed, students can enter equations into the equation pane.  Andes lacks a structured equation editor; instead, equations can be entered using conventional syntax ($=$, $+$, $-$, $*$, $/$, \^{}, and \_).  In correspondence with the second design principle, variables and equations may be entered in any order, so long as all variables included in an equation are defined before the equation itself.  Once all relevant variables and equations are defined, and numerical values for the variables are entered, a calculator finds the numerical solution to the problem \cite{gv2000}.  

The capacity to provide students with different types of feedback is one of Andes' unique characteristics.  In accordance with the third design principle, Andes provides immediate feedback once a student completes an action: if the entry is correct, it is colored green, and if it is incorrect, it is colored red.  When student mistakes are likely due to lack of attention, Andes provides unsolicited help in the form of an error message.  Common errors of this sort include leaving blank entries in dialog boxes, using undefined variables in equations, or forgetting the units of a dimensional number.  On the other hand, if an error is not recognized as this type of common mistake, Andes simply colors the entry red.  Once an entry is marked red, students can either correct their mistake without receiving any help, or they can select the entry and click on a help button, which automatically offers the student a relevant hint.  Hints are generally available in sequences of three: a pointing hint, a teaching hint, and a bottom-out hint.  These hints tend to be short, and, as suggested by the fourth design principle, are designed to point students to the feature of the entry that was incorrect.  Specifically, the pointing hint simply calls attention to the location of the student's error, so that if the student has the relevant knowledge, the mistake can be easily corrected.  However, if the student gets stuck, the teaching hints provide students with a relevant piece of knowledge that could be used toward a solution.  Finally, the bottom-out hint tells the student exactly what to change.  In order to encourage students to use these hint sequences when stuck in a problem solution, points are only subtracted when students ask for the bottom-out hint \cite{van2005}.  

\subsection{Analysis}

An extensive body of literature related to physics problem solving has accumulated over the past decade.  Though much of the work in this area had studied mathematical problem solving, problem solving in physics has been investigated through a number of careful studies comparing the problem-solving strategies of novices and experts.  The literature in this area now points to three fundamental differences between novice and expert problem solvers.  First, experts possess highly coherent, hierarchical knowledge structures that are organized around a small number of fundamental physics principles, allowing experts to efficiently evaluate the consequences of different problem-solving decisions.  Novice problem solvers, on the other hand, possess disorganized, incoherent knowledge structures of loosely connected facts and formulas.  Though important in problem solving, these knowledge structures generally become more organized as students develop more profound conceptual understanding of physics.  Second, expert physicists employ explicit strategies during problem solving, beginning with an initial problem analysis in multiple representations, and proceeding with extensive qualitative reasoning before commencing the algebraic and numerical manipulation of the problem.  Novices generally lack these strategies altogether.  Finally, metacognition plays a fundamental role in expert problem solving; experts spend a considerable amount of time analyzing problems, planning general solutions, selecting strategies, and evaluating the outcomes of their actions.  This type of metacognitive activity is deficient in most novices \cite{lar1980,dhi1996,sny2000}.  

A number of different researchers have attempted to utilize these results to create physics problem-solving curricula or instructional methods, since university physics courses generally lack explicit problem-solving instruction.  At the University of Minnesota, for instance, Heller \& Heller \cite{hh1999} have developed the Cooperative Problem Solving curriculum, in which students are required to use structured problem-solving frameworks simulating experts' structured approaches to physics problems.  These are combined with so-called context-rich problems and collaborative groups to create a complete problem-solving curriculum.  A number of publications have reported significant improvement in problem-solving ability in the students participating in this curriculum \cite{red2003,hel1999}.  In mathematics education, research carried out by Schoenfeld \cite{sch1983} has pointed out that the failure of attempted problem-solving instruction is often a result of the failure of the instruction to help students develop the metacognitive skills necessary to effectively apply the explicit problem-solving strategies they are taught.  Thereby, Schoenfeld developed an instructional method whereby students were asked a series of questions throughout their problem solving that challenged them to recognize the importance of planning, monitoring, and evaluation throughout the process.  Students' metacognitive awareness was greatly improved through this instructional method, as were their problem-solving skills \cite{red2003}.

Andes directly addresses one of the differences between novice and expert problem solvers: the structured problem-solving strategies.  When utilizing Andes, students effectively simulate the structured strategies of expert problem solvers.  Rather than jumping into the algebraic and numerical manipulation of the problem, as novices often do, students focus on its qualitative analysis: they create and label diagrams, identify known and unknown variables, and determine the physical formulas necessary to solve the problem.  Then, the numerical solution is handled automatically by Andes.  Attempts at teaching students the explicit problem-solving strategies of physics experts have been made in other curricula, and these have generally reported great success at improving students' problem-solving skills.  

The difference between previous attempts at teaching structured problem-solving strategies and the Andes ITS is the immediate feedback Andes provides at each step of the process.  According to the Andes creators, the immediate feedback provided by the interface is intended to prevent students from wasting time pursuing incorrect approaches toward a problem solution while providing them with a number of opportunities for learning \cite{gv2000}.  This objective is structurally supported both by providing students with the choice of whether or not to receive a hint from Andes, and by the brief but substantive nature of these hints.  However, the metacognitive aspect of problem solving is largely lacking from Andes' interface.  Students are not explicitly called upon to plan their solution, monitor their progress, or evaluate their decisions; these processes are all managed by Andes.  Furthermore, many expert physicists perceive problem solving as an inherently cyclical process, in which errors are both natural and necessary by forcing the problem solver to analyze and evaluate their answers and the problem-solving decisions leading up to it \cite{kuo2004}.  Andes eliminates the opportunity for students to pursue these wrong paths, thereby also eliminating very important accompanying metacognitive processes.

\subsection{Outlook}

The Andes Intelligent Physics Tutor represents an important step toward a problem-solving curriculum.  Although numerous studies have investigated the nature of physics problem solving, and a detailed picture of the cognitive structures and strategies underlying expert problem solving is emerging, the vast majority of introductory university physics courses lack explicit problem-solving instruction.  Andes, on the other hand, provides students with helpful structure and guidance in how to perform a qualitative problem analysis, and through its hints, offers plenty of opportunities for student learning.  Like Physlets, Andes adheres well to collaborative learning principles, since the ITS functions as an instructor or a peer.  Furthermore, Andes is available for free online, making it easily accessible, and though the interface may initially appear confusing, convenient videos are available online to guide students in how to interact with the tutor.  
	
However, research in physics problem solving has highlighted the importance of metacognitive activities such as planning, monitoring, and evaluation throughout the problem-solving process, and Andes does not provide students with explicit opportunities to practice these metacognitive activities.  Therefore, though Andes has the potential to play an important role in a future problem-solving curriculum, it would need to be supplemented with extensive opportunities for students to solve problems without its guidance, and with explicit direction in how to increase metacognitive awareness.

\section{Microcomputer-Based Laboratory Tools}

\subsection{Technical Specifications}

In the late 1980's, inspired by cognitive science and education research emphasizing the importance of grounding scientific principles in students' concrete experiences, the Center for Science and Mathematics Teaching at Tufts University embarked upon the ÒTools for Scientific ThinkingÓ project \cite{tho2006}.  The efforts of the project produced so-called microcomputer-based laboratory (MBL) tools and accompanying software around the central objective of helping students recognize the connections between the physical world and the abstract physics principles presented in the classroom.  These MBL tools made use of inexpensive probes --- made by Vernier Software \& Technology --- capable of measuring physical quantities such as position, velocity, force, temperature, current, and voltage, connected to the basic Apple computers of the time \cite{ver2006}.  Though computer technologies have changed significantly in the past two decades, the technical details of today's MBL tools are fundamentally the same as the 1980's versions: analog-to-digital converters (ADCs) connect to computers, and a variety of probes measuring a wide range of physical quantities are connected to the ADCs.  The accompanying software then allows for real-time graphical plotting of measured and inferred physical quantities \cite{red2003,ts1990}. 

The MBL tools were developed in order to overcome a number of technical obstacles often facing students in the traditional physics laboratory.  First, the tools can relieve students of the time-consuming and distracting traditional process of data collection and display.  Second, data is presented graphically in real time, allowing students to quickly view the data in understandable form and evaluate it.  Third, the speed with which data is collected and displayed allows students to examine a larger number of physical phenomena each laboratory period.  And, rather than spending the laboratory period collecting and plotting data, students can instead dedicate their time to analyzing and discussing the collected data.  Fourth, the general nature of the hardware and software allows these to be used to investigate a wide range of physical phenomena without having students spend a significant amount of time learning how to use complicated tools.  Finally, another consequence of the tools' general nature is that they can be used in physics classrooms of all levels, from elementary school to the university \cite{ts1990}.  

\subsection{Analysis}

In order to evaluate the pedagogical significance of MBL tools, two fundamental questions need to be addressed: first, what are the goals of the physics laboratory?  Second, how successful have traditional physics laboratories been in achieving these goals?

A wide range of goals have generally been cited for the physics laboratory.  As outlined recently by Redish \cite{red2003}, the laboratory is a place where theoretical principles and results presented in class can be verified, while providing students with mechanical skills at handling common physics apparatuses and experience with different measuring tools.  It also familiarizes students with the process of error analysis and statistics.  Furthermore, and perhaps more profoundly, laboratories can aid students in building an understanding of physics concepts, the empirical basis of science, and the nature of scientific exploration and research.  Finally, it can help students understand the importance of independent thought and coherence in scientific thinking.  

Unfortunately, most contemporary physics laboratories only address the first few goals, with occasional attention paid to error analysis.  Laboratory manuals tend to provide students with highly explicit instructions of lengthy and detailed procedures, focused on data collection and graphing, and only occasionally calling for brief responses to questions.  Studies videotaping student activity in physics laboratories have shown that students spend most of the time reading the laboratory manual, with most discussion centered on concrete questions of how to prepare the apparatus and collect data.  The process is highly prescriptive, and rarely calls for conceptual reasoning or understanding from the student.  These ÒcookbookÓ laboratories --- often accused of turning students into technicians rather than scientists --- are common in most university science classes, and few attempts have been made at altering their structure \cite{red2003,leo1991,pk2005}.  

Against this backdrop, the MBL tools enable a number of profound changes to the physics laboratory.  Theoretical principles and results from class can still be reproduced and verified.  While relieving students of the drudgery of data collection, MBL tools provide students with experience in experimental setup, though in this case they utilize the computer for the collection, display, and analysis of data.  Furthermore, students are able to practice laboratory skills such as experimental setup and calibration.  Thereby, though the experimental setups are different in the traditional laboratories and those enhanced with MBL tools, they both address the first several objectives of the physics laboratory.  One recognized disadvantage of the MBLs, however, is that at least in their current implementation, error analysis receives little attention \cite{red2003}.

It is really in the later, and more profound, objectives of the physics laboratory that the MBL tools offer a significant advantage over the traditional laboratory setup.  Unlike the traditional setup, MBL tools enable an immediate connection between physical phenomena and useful abstractions, such as the symbolic representation of physical phenomena in graphical form.  The subsequent analysis --- which can occur immediately after the phenomenon is observed --- of these representations allows theoretical concepts presented in class to be grounded in students' experiences in the laboratory.  Several studies have, in fact, reported significantly improved understanding of physics concepts in students utilizing MBL tools in their laboratories \cite{ts1990,rss1997,tg2000}.  

Furthermore, as less time is spent carrying out data collection and display, more time is available for students to experience scientific exploration and inquiry.  Rather than focusing the laboratory period on data collection, students can spend their time making predictions and hypotheses, designing experiments to test these hypotheses, and evaluating their results.  Unlike traditional laboratories where time generally allows for only one such sequence, students can modify their hypotheses, alter their experimental setups, and engage in extensive discussion with their peers.  This type of experience more closely simulates the nature of actual scientific research \cite{ts1990}.  

Finally, unlike the prescriptive nature of traditional physics laboratories, those incorporating MBL tools have a more substantial engagement/discovery component; the students thereby have more opportunities to construct their own understanding of physical phenomena and scientific principles, in accordance with modern constructivist theories of learning \cite{rss1997}.  

\subsection{Outlook}

Traditional university physics courses have generally consisted of two components: lectures and laboratories.  In formal lectures, professors present new physical principles, along with relevant derivations and, perhaps, a number of example problems.  As a result of the non-interactive nature of these lectures, laboratories have historically played an important role in the introductory physics curriculum.  The laboratory has represented the space where students can interact with teaching assistants, confirm and validate the principles and results presented during lecture, familiarize with the fundamentals of scientific research, and solidify their understanding of fundamental physics principles.  

Though numerous studies have demonstrated the failure of the traditional physics laboratory to address these goals to a satisfactory level, research in cognitive science, physics education, and educational technologies are concurrently producing instructional tools and methods that are, directly or indirectly, altering the role of the laboratory in the physics curriculum.  Class meetings are no longer guaranteed to be occupied by professor lectures and passive students; instead, interactive curricula, such as Mazur's \emph{Peer Instruction} \cite{maz1997}, are turning the classroom into a space where student misconceptions about physics are addressed directly, where students build their conceptual understanding through extensive discussion, and where students are actively involved in the construction of their physics knowledge.  Additionally, online simulations, such as Christian and Belloni's (2004) Physlets, are extending the space where students can explore physical concepts beyond the classroom and into computer labs and dormitories.  

The role of the laboratory, then, is fundamentally altered.  In fact, some introductory physics courses, such as Physics by Inquiry at the University of Washington \cite{mcd2006} and Workshop Physics at Dickinson College \cite{law2004}, are abolishing the division between lecture and laboratory altogether, teaching the course in a so-called workshop or studio method.  In these classes, the lecture plays only a small or nonexistent role, and instead, students spend the period working with laboratory equipment in collaborative groups.  In the Workshop Physics curriculum, heavy use is also made of MBL tools.  In light of our current understanding of the importance of constructivist classrooms and collaborative learning, these examples of completely revolutionized classrooms represent a tremendous step ahead in contemporary physics education.

Given their demonstrated effectiveness in helping students develop a more profound conceptual understanding of physics principles, MBL tools hold great promise in improving introductory university physics curricula.  Coupled with curricula grounded in firmly established theories of learning --- such as constructivism, cognitive conflict in conceptual change, and social learning --- MBL tools can bring some of the most profound objectives of the traditional laboratory into reality in the modern physics classroom.  

\section{Conclusion}

The emergence of educational technologies in physics education is having a profound impact on all areas of physics instruction, from course management to problem-solving instruction and to data collection in the laboratory.  Although it does not offer a panacea to the wide range of problems plaguing physics education, the emergence of these technologies is forcing cognitive scientists, physics education researchers, physics instructors, and educational technology researchers to critically evaluate the goals of physics education and our understanding of how students learn physics.  Only through this important interdisciplinary collaboration can physics instruction be fundamentally improved.

The technologies presented and discussed in this paper offer a glimpse into the tremendous potential of educational technologies in physics education.  Physlets allow students to visualize and interact with highly complex and abstract physical concepts; Andes guides students through expert-like approaches to physics problems; MBL tools relieve students of the drudgery of data collection in the laboratory and helps them establish the connections between physical phenomena and abstract representations.  Physics education researchers have already created numerous structured curricula founded on well-established principles of learning --- such as constructivism and social learning --- that integrate these technologies.  Though implementing these curricula will be a tremendous challenge, physics students will surely benefit from the gradual elimination of the non-interactive, passive lecture as the basis of university physics education.

\end{document}